\documentclass[aps,prd,nofootinbib,preprint,superscriptaddress]{revtex4-1}
\usepackage{amssymb}
\usepackage{amsmath}
\usepackage{amsfonts}
\usepackage{graphicx,color}
\usepackage{mathrsfs}
\usepackage{hyperref}

\def\be{\begin{equation}}
\def\ee{\end{equation}}
\def\ber{\begin{eqnarray}}
\def\eer{\end{eqnarray}}
\def\bwt{\begin{widetext}}
\def\ewt{\end{widetext}}

\begin{document}

\author{Sujoy K. Modak}
\email{smodak@ucol.mx}
\affiliation{Facultad de Ciencias - CUICBAS, Universidad de Colima, Colima, C.P. 28045, M\'exico}
\affiliation{KEK Theory Center, High Energy Accelerator Research Organization (KEK),\\ Tsukuba, Ibaraki 305-0801, Japan}


\title{New geometric and field theoretic aspects of a radiation dominated universe}

\begin{abstract}
The homogeneous and isotropic radiation dominated universe, following the inflationary stage, is expressed  as a spherically symmetric and inhomogeneous spacetime upon a power law type conformal transformation of the null (cosmological) coordinates. This new spacetime metric has many interesting properties. While the static observers, at a fixed position in this new spacetime, do not see any horizon, some non-static observers encounter a horizon due to their motion which is analogous to the situation of Rindler observers in Minkowski spacetime. The symmetry of the new metric offers a unitarily inequivalent quantisation of the massless scalar field and provides a new example of particle creation. We calculate the particle content of the cosmological vacuum state with respect to the static observer in this new spacetime who, with respect to cosmological time, is freely falling in asymptotic past and future but accelerated in between.
\end{abstract}

\maketitle
\tableofcontents

\section{Introduction}
The discovery of Hawking that black holes radiate \cite{hawk1}, and,  the mathematical demonstration of it \cite{hawk2} which depends on the unitarily inequivalent quantisation of matter fields in the infinite past and future,  had initiated the subject of particle creation in curved space of considerable interest in last few decades. Another seminal work by Unruh \cite{unruh} has made this even more interesting by showing another example of particle creation for accelerating observers in flat space. Other interesting examples of particle creation in curved space include the expanding universe (such as when the initial Bunch-Davies vacuum state gets excited due to cosmic expansion \cite{Parker:1968mv}-\cite{Aoki}, particle creation in the static de-Sitter universe \cite{Mott, Singh:2013pxf}), moving mirror set up \cite{fulling, davies} etc.

One of the features that is common, in most of the above situations, is the existence of at least two reference frames (existing simultaneously or causally  separated), non-inertially related with each other, allowing systematic field quantisations so that there are more than one way to decompose the field operator into positive and negative frequency modes. In fact, one of the salient features of particle creation in curved space is the close inter-dependency of the geometry and field theoretic aspect of the quantum fields, living in that geometry. Given a favourable situation, such as allowing unitarily inequivalent  quantisations, inherits a property that  the positive frequency mode of one of these frames is necessarily a superposition of the positive and negative frequency modes in the other frame, resulting  a non-vanishing set of Bogolyubov coefficients connecting field modes/operators, which, then leads to the non-vanishing particle number density. We refer the interested first time readers to the following (incomplete) list of review articles \cite{DeWitt:1975ys}-\cite{Brandenberger:1984cz} and monographs \cite{birrel}-\cite{parker} on these aspects.

In this paper we introduce a new set of coordinates to describe the radiation dominated epoch of the expanding universe. The new spacetime metric describing this epoch, which is found by making a conformal transformation of the cosmological spacetime, is a spherically symmetric, inhomogeneous spacetime. Any static observer, at a fixed position, in this spacetime, has an interesting dynamics in terms of cosmological coordinates - in asymptotic past and future the static observer is indistinguishable from the freely falling observer but in between it has both acceleration and deceleration. Further, some non-static observers, unlike others, cannot access the full spacetime and encounters a horizon whose location depends on its motion. This is qualitatively very similar to the Rindler observers, who have a constant acceleration in Minkowski spacetime and where they encounter the Rindler horizon just because of their accelerated motion. Like Rindler coordinates which only covers one-fourth of the full Minkowski spacetime, some non-static frames here also do not cover the full spacetime accessible in cosmological frame.

We further report a new example of particle creation in the radiation dominated (expansion) phase of the FRW universe. We quantise the massless scalar field (which may also be conformally coupled since the Ricci curvature is zero for radiation epoch) in cosmological coordinates and new coordinates that we introduce here. Then we calculate the particle content of the cosmological vacuum state with respect to the static observer in the new frame, both, in two and four spacetime dimensions.

The paper is organised as follows. In Section II, we make a conformal transformation of the FRW metric with cosmological coordinates. In Section III, we discuss the interesting and special situation exhibited by the radiation dominated universe. In section IV we discuss the static versus non-static observer trajectories and appearance of a new horizon for certain non-static observers. Section V includes the discussion of the particle creation in a two dimensional situation. In Section VI, we discuss the four dimensional set up of particle creation. Finally, Section VII includes discussions and future outlook. 


\section{Conformal transformation of cosmological coordinates }
We start with the spatially flat FRW metric
\begin{eqnarray}
ds^2= dt^2 - a^2(t)[dr^2 + r^2(d\theta^2 +{\sin^2{\theta}}~ d\phi^2)].
\label{frw}
\end{eqnarray}
after writing it in a conformally flat form using the cosmological time $\eta = \int \frac{dt}{a(t)}$, we get
\begin{eqnarray}
ds^2=a^2(\eta)[d\eta^2 - dr^2 - r^2(d\theta^2 +\sin^2{\theta}~ d\phi^2)].
\label{cffrw}
\end{eqnarray}
 The conformally flat metric in the light-cone gauge, using $u = \eta-r$, $v=\eta+r$ and $r=\frac{v-u}{2}$ becomes
\begin{equation}
ds^2 = a^2 du dv - \frac{(v-u)^2}{4}a^2 (d\theta^2 + \sin^2\theta d\phi^2).\label{ncf}
\end{equation}

Now we make a new, power law type coordinate transformation (of the conformal type)
\begin{eqnarray}
V= \frac{1}{\lambda m} v^m \label{Vv}\\
U = \frac{1}{\lambda m} u^m\label{Uu}
\end{eqnarray}
for $u>0$ (with a constant $\lambda$) and,
\begin{eqnarray}
V= \frac{1}{\lambda m} v^m \label{Vv}\\
U = -\frac{1}{\lambda m} u^m\label{Uu}
\end{eqnarray}
for $u<0$, respectively. 

To express \eqref{ncf} in $U,V$ coordinates, we also need to transform the scale factor. First we note  that $a (t)=a_0 t^n$, so that $\eta = \int\frac{dt}{a(t)} = \frac{t^{1-n}}{a_0 (1-n)}$, implying $a (\eta)=a_0 (a_0 (1-n)\eta)^{\frac{n}{1-n}}$. Then by using $\eta=\frac{u+v}{2}$ and substituting $u$ and $v$ from \eqref{Uu} and \eqref{Vv}, respectively, we can obtain $a(U,V)$. The final result of the metric \eqref{ncf} in new ($U,V$) null coordinates is,
\begin{eqnarray}
ds^2 = A (U,V) dU dV - B(U,V) (d\theta^2 + \sin^2\theta d\phi^2),
\label{newUV}
\end{eqnarray} 
where 
\bwt
\begin{eqnarray}
A (U,V) &=&  (\lambda^2 a_0^2) (\pm \lambda^2  m^2 U V)^{\frac{1}{m}-1}  [a_0(1-n)/2]^{\frac{2n}{1-n}}    \left(  (\pm \lambda m U)^{1/m}+(\lambda  m V)^{1/m}\right)^{\frac{2n}{1-n}} \label{A} \\
B (U,V) &=& \frac{a_0^2}{4} [a_0(1-n)/2]^{\frac{2n}{1-n}} \left( (\lambda  m V)^{1/m} - (\pm \lambda m U)^{1/m}\right)^2 \left(  (\pm \lambda m U)^{1/m}+(\lambda  m V)^{1/m}\right)^{\frac{2n}{1-n}} \label{B}  \nonumber \\
\end{eqnarray}
\ewt
For arbitrary values of $m,n$, the above interval is of course very nontrivial and is not associated with any symmetry. However, there is one exception, as for the case $m=2$ and $n=1/2$, i.e., for the radiation dominated universe which we will discuss in the next section.

\section{Spherically symmetric form of radiation dominated universe}

Clearly, for radiation dominated phase ($m=2$ and $n=1/2$) \eqref{A} and \eqref{B} becomes 
\begin{eqnarray}
A (U,V) &=& \left(\frac{\lambda^2 a_0^4}{4}\right) \frac{(\sqrt{V} \pm \sqrt{\pm U})^2}{4 \sqrt{\pm UV}}\\
B (U,V) &=& \left(\frac{\lambda^2 a_0^4}{4}\right) \left(\frac{V- (\pm U)}{2}\right)^2
\end{eqnarray}
and the metric \eqref{newUV} takes the form
\begin{equation}
ds^2 = \frac{\lambda^2 a_0^4}{4} d\tilde{s}^2
\end{equation}
where
\begin{equation}
d\tilde{s}^2 =  F(U,V) dU dV -    G(U,V) d\Omega^2 \label{newrd}
\end{equation}
with
\begin{eqnarray}
F(U,V) = \frac{(\sqrt{V} \pm \sqrt{\pm U})^2}{4\sqrt{\pm UV}},\\
G(U,V) = \left(\frac{V- (\pm U)}{2}\right)^2.
\end{eqnarray}
Once again `$+$' and `$-$' signs are applicable for $U>0$ or $U<0$, respectively. Demanding $ds^2 = d\tilde{s}^2$ implies $\frac{\lambda^2 a_0^4}{4} =1$. For a universe transiting to radiation stage  from the inflationary stage one can easily calculate (by equating the scale factor and its derivative at the transition point) $a_0=\sqrt{2{\cal H}e}$ (see \cite{Singh:2013bsa}) which implies
\be
\lambda = 1/{\cal H}e \label{lmda}
\ee
where ${\cal H}$ is the Hubble constant of the inflationary stage of the universe. This, therefore imply the following relationships between two null coordinates
\be
U=\pm \frac{{\cal H}e}{2}u^2 ; \hspace{1cm}
V=\frac{{\cal H}e}{2}v^2 \label{Uu-n}
\ee
where again $+(-)$ sign stands for $u>0 (u<0)$.

At this point we introduce the new time and radial coordinates
\begin{eqnarray}
T = (V+U)/2\label{RTN2}\hspace{1cm}; \hspace{1cm} R= (V-U)/2. \label{RTN1}
\end{eqnarray}
In these coordinates the spacetime \eqref{newrd} for $U \ge 0$ or $T\ge R$ (Region-I) becomes
\be
d{s}_I^2 = F_I(T,R) (dT^2 -dR^2) - R^2 d\Omega^2 \label{mRT00}
\ee
with
\be
F_I(T,R) = \frac{(\sqrt{T+R} + \sqrt{T-R})^2}{4\sqrt{T^2 - R^2}}. \label{FTR1}
\ee
Whereas, for $U \le 0$ or $T\le R$ (Region-II) we get
\be
d{s}_{II}^2 = F_{II}(T,R) (dT^2 -dR^2) - T^2 d\Omega^2 \label{mRT11}
\ee
with
\be
F_{II}(T,R) = \frac{(\sqrt{R+T} - \sqrt{R-T})^2}{4\sqrt{R^2 - T^2}}. \label{FTR2}
\ee
 
In region I (\eqref{mRT00}) 
\ber
T &=& (V+U)/2 = \frac{1}{2\lambda}(\eta^2 + r^2) \label{RT2}\\
R &=& (V-U)/2 =\frac{\eta r}{\lambda} \label{RT1}.
\eer
Also, for region II, the relationships between the two sets of coordinates are reversed, so that
\ber
T &=& (V+U)/2 =\frac{\eta r}{\lambda} \label{ter}  \\
R &=& (V-U)/2 =  \frac{1}{2\lambda}(\eta^2 + r^2) \label{uf1}\label{rer}.
\eer

We can also express the conformal factor $F(T,R)\rightarrow F(H(T,R),R)$ and express the intervals \eqref{mRT00} and \eqref{mRT11} in more convenient forms, by introducing the Hubble parameter. We do it separately for regions I and II.

Using \eqref{RT2}, \eqref{RT1} and the Hubble parameter $H(\eta) = \frac{2}{a_0^2 \eta^2}$ (and noting $\frac{\lambda^2 a_0^4}{4} =1$), we obtain 
\be
1-H^2R^2 =  1- \frac{ r^2}{\eta^2}. \label{uf3}
\ee
Since in region I $u\ge 0$ or $\eta\ge r$, we must have, as follows from the above equation, $R\le 1/H$.  From \eqref{RT2} and \eqref{RT1} we can substitute $\eta,r$ in \eqref{uf3} to obtain
\be
1-H^2R^2= \frac{1}{F_I(T,R)}
\ee
so that (\eqref{mRT00}) can also be expressed as
 \bwt
 \ber
ds_I^2 &=& \frac{dT^2 - dR^2}{1-H^2R^2} -R^2 (d\theta^2 + \sin^2\theta d\phi^2) ,~~\text{for}~~R\le1/H. \label{mRT2}
\eer
\ewt
For region II, it is easy to check
\ber
1-H^2T^2= 1- \frac{ r^2}{\eta^2} = - \frac{1}{F_{II}(T,R)}.
\eer
Since, in region II $u\le 0$ or $\eta\le r$, the above equation states that, in this region we must have $T\ge 1/H \implies H^2T^2 -1\ge 0$ (note that it also means $R\ge 1/H$, since, $R>T$ in region II)
so that
\bwt
\ber
ds_{II}^2       &=& \frac{dT^2 - dR^2}{H^2T^2 -1} -T^2 (d\theta^2 + \sin^2\theta d\phi^2),~~\text{for}~~R\ge1/H, \label{mRT3}
\eer
\ewt
which is slightly different than \eqref{mRT2}. Here, the radius of the two sphere is timelike while for \eqref{mRT2} the radius is spacelike. It is also clear that $T$ is always timelike and $R$ is always spacelike since we do not encounter any signature change between \eqref{mRT2} and \eqref{mRT3}, while passing from sub to super Hubble scale. To the best of our knowledge, these spacetime metrics, representing the radiation stage of the universe, is a new input in this paper. We invite the interested reader to study further the various other usage of these metrics, elsewhere.

Now let us have a closer look on the various interesting properties of the new spacetime metrics, take for instance, the forms, as given in \eqref{mRT2} and \eqref{mRT3} . The entire spacetime which was originally defined by the cosmological (null) coordinates for $ 0< \eta<\infty$ and $0 <r <\infty$ in \eqref{cffrw} is covered by two new sets of coordinates $0 <T < \infty$ and $0<R<\infty$, jointly by \eqref{mRT2} and \eqref{mRT3}  (or equivalently by \eqref{mRT00} and \eqref{mRT11}). Unlike the cosmological case where, $H$ is only a function of cosmological time $\eta$ (or comoving time $t$), here $H$, expectedly  is a function of both $T$ and $R$. The universe looks different for different observers, situated at various spatial points of the universe and, the Hubble parameter is not constant for a given $T$ for these observers (with $T$ as their proper time). The new metrics,  \eqref{mRT2}, \eqref{mRT3},  are although isotropic (because of the spherical symmetry),  they are inhomogeneous - the metric coefficients not only depend on $T$ but also on $R$. This is in contrast to the case where all comoving observers all share the same Hubble value at any given time. For an observer at small $R<<1/H$, one may ignore the quadratic term as compared to 1 in \eqref{mRT2} and then the metric becomes homogeneous for all observer in the vicinity. However, as one approaches the  Hubble scale $R\sim1/H$ the metric becomes highly inhomogeneous{\footnote{The metric \eqref{mRT3}, describing the super-Hubble region, is always inhomogeneous except when $T$ approaches zero.}}. The physical reason is rooted in the conformal transformation -  it does not respect the homogeneity for the radiation case.

\section{Static vs non-static observers, and horizons}

In this section we argue that, in the new ($T,R,\theta,\phi$) coordinates, static observers at $R=const.$ and other non-static observers with some motion encounter very different physical reality. In particular, while the static observer can get the information, by respecting causality, from anywhere in the spacetime, some  non-static observers encounter a horizon. The occurrance of this horizon with respect to the non-static observers is qualitatively similar to the Rindler horizon, in a sense, that the latter is only experienced  by accelerated  observers in flat/Minkowski spacetime and not by inertial observers. 

\begin{figure}[t]
\centering
\includegraphics[angle=0,width=10cm,keepaspectratio]{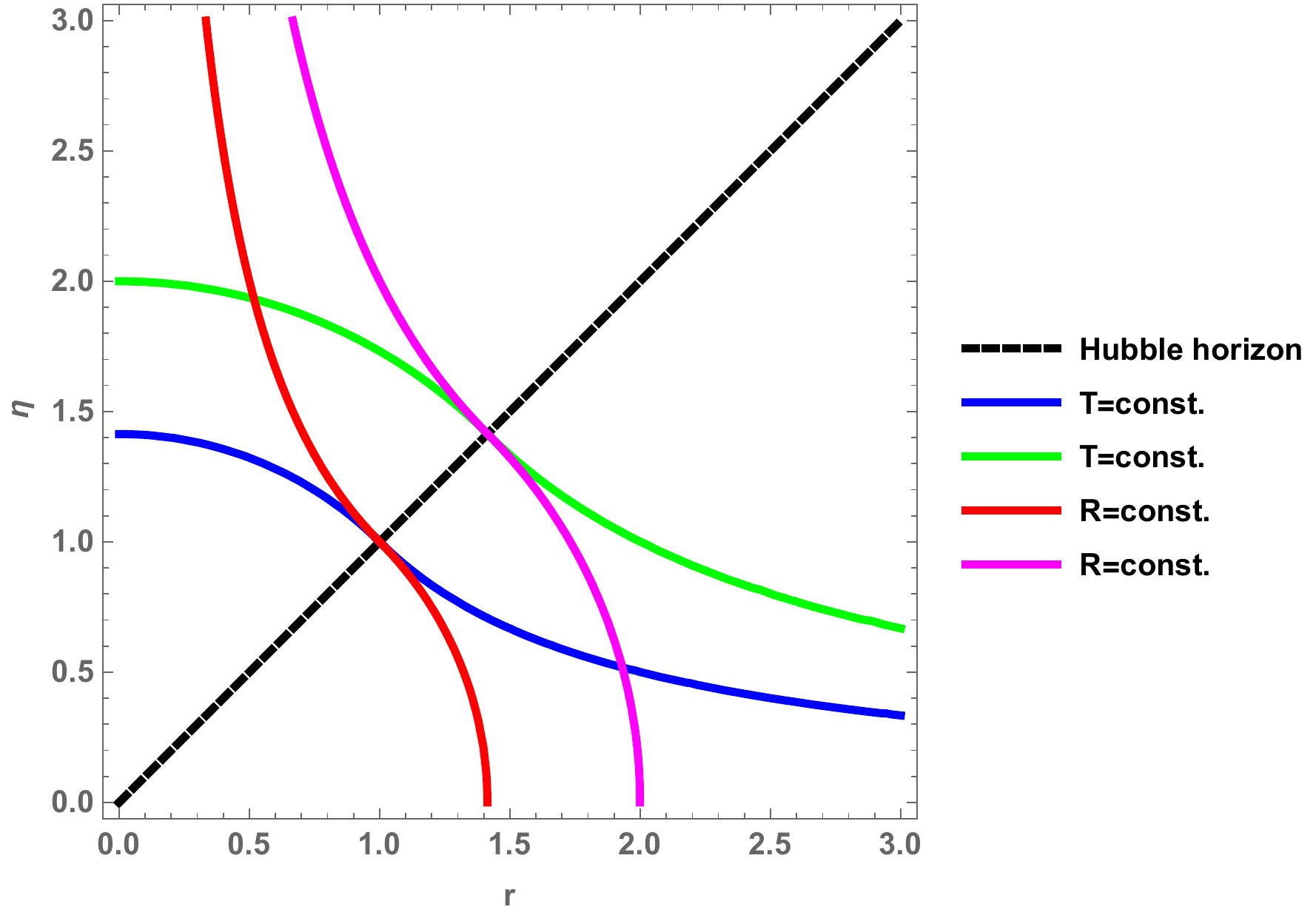}
\caption[]{$T$ and $R$ constant curves in ($\eta,~r$) plane. The intersection points are on the light-cone boundary at Hubble scale. Note that $T$ always remain timelike, $R$ always remain spacelike. Any static observer at $R=const.$ can get causal information from any point in the full spacetime. There is no horizon for a static observer at the Hubble scale. However, non-static observers find a horizon.}
\label{fig1}
\end{figure}

First, let us plot the $T=const.$ and $R=const.$ orbits in the $(\eta,~r)$ plane in Fig. \ref{fig1}, for regions I and II.  From \eqref{RT2} and \eqref{RT1}, it is clear that the $T=const.$ and $R = const.$ curves in the ($\eta,~r$) plane are, respectively, given by the circular and linearly inverse trajectories in region-I, while, these are just opposite  in region-II. Notice, that any $R=const.$ trajectory in asymptotic past ($\eta \rightarrow 0$) and in asymptotic future ($\eta \rightarrow \infty$) are indistinguishable from the freely falling observers. As time progresses the inertial observer, defined at infinite past, starts accelerating and attains the luminal velocity near the Hubble scale. It crosses the Hubble radius, from super to sub region, with the speed of light. After that, the observer starts decelerating (in an identical rate of acceleration at super Hubble scale) and finally becomes freely falling towards the centre ($r=0$), at asymptotic future. Note that, this static observer can obtain signal from any where in the spacetime - there is no horizon for her. This fact is also reflected in the fact there is no signature change in the metric \eqref{mRT00} and \eqref{mRT11}. The static observer while approaching the horizon, as evident from Fig. \ref{fig1}, follows the trajectory $\eta = -r + \text{const}$ (or $T=-R + \text{const}$) for a moment, and along this line although $F_I$ and $F_{II}$ diverges, we also have $dT=-dR$ and, therefore, the factor multiplying $F_I$ and $F_{II}$ vanishes, which makes the interval $ds^2$ to vanish.

Now let us consider the case of a non-static observer following a trajectory $T= G(R)$ so that $dT^2-dR^2= (G'^2(R)-1)dR^2$ and the $T,R$ sector of \eqref{mRT00} becomes
$$\frac{[\sqrt{G(R)+R} + \sqrt{G(R)-R}]^2}{4\sqrt{G(R)^2 - R^2}}(G'^2(R)-1)dR^2.$$
Clearly, for any root $R_0$ satisfying - (i) $G(R_0)-R_0 = 0$ and, (ii) $G'(R_0) \ne \pm 1$, the interval diverges at $T= G(R_0)$. This observer will encounter a horizon. Let us consider the example of the trajectory $T=G(R)=\alpha R^2$, where $\alpha$ is a dimension-full, positive definite constant. In region I, using the relationships \eqref{RT2} and \eqref{RT1} and the expression for $\lambda$, this trajectory $T=G(R)=\alpha R^2$ becomes
\be
\eta^2 = \frac{1}{2\alpha{\cal H}e} \left(\frac{r^2}{r^2 - \frac{1}{2\alpha {\cal H}e}}\right).
\label{ns1}
\ee
Since the right hand side must satisfy the condition of positive definiteness, we must have $r\geq\frac{1}{\sqrt{2\alpha {\cal H}e}}$. This clearly implies that, this non-static observer will not have access to the region of spacetime $0<r<\frac{1}{\sqrt{2\alpha {\cal H}e}}$, i.e., she will encounter a horizon at $r = \frac{1}{\sqrt{2\alpha {\cal H}e}}$ which appears only due to her motion. This is strikingly similar for Rindler observers who encounters a horizon due to their own acceleration. The positions of the horizon for this observer is given by the roots of $R_0 - \alpha R_0^2 = 0$, which are  $R_0=0$ (the inner horizon) and $R_0 = 1/{\alpha}$  (the outer horizon). If we inspect them physically then we need to discard $R_0=0$ (which can be reached by setting $r=0$) from the physical patch because this will be inaccessible to the physical spacetime that the non-static observer sees, for which, as follows from \eqref{ns1}, $r\geq\frac{1}{\sqrt{2\alpha {\cal H}e}}$. The position of the horizon $R_0 = 1/{\alpha}$, on the other hand, corresponds to $r = \frac{1}{\sqrt{2\alpha {\cal H}e}}$.

Coming back to the static case where $G(R) = -R + \text{const}$, from an analogous relationship of \eqref{ns1}, between $\eta$ and $r$, one can easily conclude that the horizon is nonexistent in this case. Further,   as depicted in Fig. \ref{fig1}, the $T=const.$ slices offer a new Cauchy slicing of the spacetime (along with $\eta=const.$ slicing). Any initial Cauchy data has a well defined time evolution through these Cauchy slices. Therefore, we see, there is a recipe of an alternative quantisation of matter fields using $T,R$ coordinates and a framework for studying particle creation. The rest of the paper is dedicated to study this particle creation process. We shall restrict ourselves to the static observers' case who will encounter the cosmological vacuum state (defined with respect to the observers with proper time $\eta$) containing particles.

\section{Particle creation: two dimensional analysis}
We first consider the two dimensional toy version of radiation dominated universe which is  simpler but useful to get a physical insight of particle creation process. Since any two dimensional spacetime can be written as a conformally flat metric, the massless Klein-Gordon equation is much simpler, which for\eqref{ncf} and \eqref{newrd} (by ignoring $\theta,~\phi$ part) just become
\begin{eqnarray}
\partial_u\partial_v \Phi= 0,\\
\partial_U\partial_V \Phi=0
\end{eqnarray}
respectively, with plane-wave mode solution, leading to following expansions of the field operator 
\ber 
 \hat{\Phi} &=& \int_o^{\infty} \frac{d\omega}{\sqrt{4\pi\omega}}(e^{-i\omega u} a_\omega + e^{i\omega u} a_\omega^{\dagger} + \text{right moving}) \nonumber\\ \label{exp1}\\
 &=& \int_o^{\infty} \frac{d\omega}{\sqrt{4\pi\Omega}}(e^{-i\Omega U} b_\Omega + e^{i\Omega U} b_\Omega^{\dagger} + \text{right moving}) \nonumber\\ \label{exp2}.
\eer 
Now we use the Bogolyubov transformation for the left moving modes (calculation for right moving modes are fully analogous and is left out here)
\be
b_\Omega = \int_0^{\infty} d\omega (\alpha_{\Omega\omega} a_\omega - \beta_{\Omega\omega} a_\omega^{\dagger})
\ee
and substitute this in \eqref{exp2} and comparing we infer
\be
\frac{1}{\sqrt{\omega}} e^{-i\omega u} = \int_0^\infty\frac{d\Omega'}{\sqrt{\Omega'}} \left(\alpha_{\Omega'\omega} e^{-i\Omega' U} - \beta_{\Omega'\omega} e^{i\Omega' U}\right).
\ee
Multiplying both sides with $e^{\pm i \Omega U}$ and integrating we obtain
\ber
\alpha_{\Omega\omega} &=& \frac{1}{2\pi}\sqrt{\frac{\Omega}{\omega}} \int_{-\infty}^{\infty} dU e^{-i\omega u +i\Omega U}, \\
\beta_{\Omega\omega} &=& -\frac{1}{2\pi}\sqrt{\frac{\Omega}{\omega}} \int_{-\infty}^{\infty} dU e^{i\omega u +i\Omega U}.
\eer
Since we have $U=\pm\frac{u^2}{2\lambda}$ ($``+ " $ for $U\ge 0$ and $``-"$ for $U\le 0$) for $\beta_{\Omega\omega}$ we find
\bwt
\ber
\beta_{\Omega\omega} &=& - \frac{1}{2\pi\lambda}\sqrt{\frac{\Omega}{\omega}} \left( \int_{0}^{\infty} (u du) e^{i\omega u + i\Omega u^2/2\lambda} + \int_{-\infty}^0 (-u du) e^{i\omega u -i\Omega u^2/2\lambda}\right),\\
&=&  I + I^* \label{bog2d}
\eer
\ewt
 The first integral can be performed as follows. We first write this in the following form
\be 
I =  - \frac{1}{2\pi\lambda}\sqrt{\frac{\Omega}{\omega}} e^{-i\frac{\lambda\omega^2}{2\Omega}}\int_0^\infty u du ~e^{\frac{i\Omega}{2\lambda}(u+\frac{\lambda\omega}{\Omega})^2}.
\ee
Then defining $x=(u+\frac{\lambda\omega}{\Omega})^2$ and substituting in the integral we get
\bwt
\be 
I = -\frac{e^{-i\lambda\omega^2/2\Omega}}{4\pi\lambda}\sqrt{\frac{\Omega}{\omega}}\left( \int_{\frac{\lambda^2\omega^2}{\Omega^2}}^\infty {dx~ e^{-\frac{\Omega x}{2i\lambda}}}  - ({\lambda\omega}/{\Omega})  \int_{\frac{\lambda^2\omega^2}{\Omega^2}}^\infty {dx ~x^{-1/2} ~e^{-\frac{\Omega x}{2i\lambda}}} \right)
\ee
\ewt
The integral of above kind is often found in the calculation of Bogolyubov coefficients \cite{paddy2} and is usually calculated by using the identity
\bwt
\be
\int_{x_0}^{\infty} dx x^{s-1} e^{-bx} = e^{-s\log b} \Gamma[s, x_0];~~ Re~ b, Re~ s >0.
\label{id}
\ee
\ewt
where $\Gamma[s, x_0]$ is an upper incomplete gamma function. Two more comments are in order. First, one puts a small real number ($\epsilon\approx 0$) as the real part of $b$ in above equation and take this to zero after performing the integration and, the second, in the calculations involving Unruh or Hawking effect one usually ends up with a complete gamma function rather than and incomplete one.\\
Using this, we find
\bwt
\be
I = - \frac{e^{-\frac{i\omega^2\lambda}{2\Omega}}}{2\pi\lambda} \sqrt{\frac{\Omega}{\omega}} \left( e^{-\log|\Omega/2i\lambda + \epsilon|}~\Gamma[1, \omega^2\lambda^2/\Omega^2] -  ({\lambda\omega}/{\Omega}) e^{-\frac{1}{2}\log|\Omega/2i\lambda + \epsilon|}~\Gamma[1/2, \lambda^2\omega^2/\Omega^2] \right).
\ee
\ewt
Now one can use the relation $\lim_{\epsilon\rightarrow 0}\log| \Omega/2i\lambda +\epsilon| \simeq \log|  \Omega/2\lambda| - \frac{i \pi}{2}$ (using $\Omega >0$) to obtain
\bwt
\be
I = - \frac{e^{-\frac{i\lambda\omega^2}{2\Omega}}}{2\pi\lambda} \sqrt{\frac{\Omega}{\omega}} \left(\frac{2\lambda}{\Omega} e^{i\pi/2}\Gamma[1, \lambda^2\omega^2/\Omega^2] - \frac{\lambda\omega}{\Omega} \sqrt{2\lambda/\Omega} e^{i\pi/4} \Gamma[1/2, \lambda^2\omega^2/\Omega^2] \right).\label{2d1}
\ee
\ewt
Therefore  using \eqref{bog2d} we get
\bwt
\be
\beta_{\Omega\omega} = \frac{\sqrt{\frac{\Omega }{\omega }} \left(\omega  \sqrt{\frac{\lambda }{\Omega }} \Gamma \left(\frac{1}{2},\frac{\lambda ^2 \omega ^2}{\Omega ^2}\right) \left(\sin \left(\frac{\lambda  \omega ^2}{2 \Omega }\right)+\cos \left(\frac{\lambda  \omega ^2}{2 \Omega }\right)\right)-2 \sin \left(\frac{\lambda  \omega ^2}{2 \Omega }\right) \Gamma \left(1,\frac{\lambda ^2 \omega ^2}{\Omega ^2}\right)\right)}{\pi  \Omega }
\ee
\ewt
Now using \eqref{lmda}  we can replace $\lambda$ with the inflationary Hubble constant. Which then gives
\bwt
\be
\beta_{\Omega\omega} = \frac{\sqrt{\frac{\Omega }{\omega }} \left(\sqrt{e} \omega  \sqrt{\frac{1}{{\cal H} \Omega }} \Gamma \left(\frac{1}{2},\frac{e^2 \omega ^2}{{\cal H}^2 \Omega ^2}\right) \left(\sin \left(\frac{e \omega ^2}{2 {\cal H} \Omega }\right)+\cos \left(\frac{e \omega ^2}{2 {\cal H} \Omega }\right)\right)-2 \sin \left(\frac{e \omega ^2}{2 {\cal H} \Omega }\right) \Gamma \left(1,\frac{e^2 \omega ^2}{{\cal H}^2 \Omega ^2}\right)\right)}{\pi  \Omega }
\label{betah}
\ee
\ewt
This is an exact expression and one can readily use to calculate the exact particle number density as defined by the first equality in \eqref{def}. However, it is difficult to find an analytical expression in closed, desirable form, for which one needs to use numerics. Here, we limit ourselves to analytical approximation. 

We can make an analytical expression, in the limiting case of  $\frac{\omega}{\Omega} << {\cal H}$, i.e., the ratio between the frequencies is much less than inflationary Hubble constant (${\cal H}\sim 10^{37} s^{-1}$). This inequality has a nice physical interpretation in the following manner - if we are counting the particle excitation of frequency $\Omega \sim {\cal O}(1)$, then this inequality will ensure that we do not sum over any field modes in the inflationary era for which $\omega \sim \omega_{inf}\sim 10^{37} s^{-1}$. Thus our approximation will hold quite strongly while counting particle excitation for higher frequency ($\Omega$) than compared to its lower values.  Under such a condition   \eqref{betah} simplifies to
\be
\beta_{\Omega\omega} \sim \sqrt{\frac{e}{\pi {\cal H}}} \frac{\sqrt{\omega}}{\Omega}
\label{betap}
\ee
and the particle number density is given by
\ber
\langle n_{\Omega} \rangle &\sim&  \int_{0}^{\omega} d\omega~|\beta_{\Omega\omega}|^2 \label{def}\\
&=& \frac{e}{2\pi {\cal H}}\left(\frac{\omega}{\Omega}\right)^2. \label{n2d}
\eer
If we now substitute $\frac{\omega}{\Omega}=\epsilon{\cal H}$ we get
\ber
\langle n_{\Omega} \rangle &\sim& \frac{ {\cal H}e}{2\pi } \epsilon^2,
\eer
which is interestingly independent of $\Omega$ and proportional to the inflationary Hubble constant which is quite large (${\cal H} \sim 10^{37} s^{-1}$). However, it is multiplied with $\epsilon^2$ which makes this quite small. It can of course give a finite contribution to the energy density if one counts all the particles during the entire lifetime of the radiation dominated universe. To go beyond this special case one must consider a numerical analysis.

\section{Particle creation: four dimensional analysis}

\subsection{Quantum fields in conformally flat FRW}
We now consider a massless scalar field with arbitrary coupling and solve the Klein-Gordon equation $\Box \Phi =0$  (Ricci scalar $R$ being zero here) in the radiation dominated universe, using the metric \eqref{cffrw}. Using the separation of variables and considering the background geometry, we express
\be 
\Phi (\eta, r, \theta, \phi) = \sum_{l} \frac{f_l (r)}{r} g(\eta) Y_{lm} (\theta, \Phi) \label{anphi}
\ee
where, the angular part $Y_{lm}$ are the spherical harmonics, and, the $(\eta,~r)$ dependent parts are the solutions of the following equations (after considering the scale factor for the radiation phase)
\ber
\eta^2 \frac{d^2 g}{d\eta^2} + 2\eta \frac{dg}{d\eta} + \omega^2 \eta^2 g  &=& 0 \label{etag}\\
\frac{d^2 f_l}{dr^2} + \left(\omega^2 - \frac{l(l+1)}{r^2} \right)f_l &=& 0. \label{rf}
\eer
We can rescale $\eta' = \omega\eta$ and express \label{etag} as a spherical Bessel equation with $n=0$
\be
\eta'^2 \frac{d^2 g}{d\eta'^2} + 2\eta' \frac{dg}{d\eta'} + \eta'^2 g  = 0
\ee
whose solution is given by the spherical Bessel function of the first kind $j_{0} (\eta') = \sin\eta'/\eta'$ and the second kind $n_0(\eta') = - \cos\eta'/\eta'$. Equivalently we can construct the linear combinations $h^{(1/2)}_0 (\eta') = -n_{0} (\eta') \pm i j_{0} (\eta')  = {e^{\pm i \eta'}}/{\eta'} =  {e^{\pm i \omega\eta}}/{\omega\eta}$ to make $\Phi$ the real scalar field. Moreover, if we consider the s-wave approximation ($l=0$ in \eqref{rf}), we have $f_0 (r) = e^{\pm \omega r}$ and the ansatz \eqref{anphi} can be expressed as 
\ber
\Phi &=& \sum_{\omega} {\cal N} \frac{f_0 (r)}{r} g_{\omega}(\eta) Y_{00} (\theta, \Phi) \\
&=& \sum_{\omega} {\cal N} \left( \frac{e^{-i\omega(\eta - r)}}{2\sqrt{\pi}~\omega \eta r} a_\omega^{L} + \frac{e^{-i\omega(\eta + r)}}{2\sqrt{\pi}~\omega \eta r} a_\omega^{R} \right) + \text{h.c} \label{clphi}
\eer
where, ${\cal N}$ is a normalization constant to be determined by imposing the orthogonality condition of the inner product between field modes; $L$ and $R$ stands for the left and right moving field modes. The coefficients $a_\omega^{L/R}$ are to be elevated to the annihilation operators for two oppositely moving modes at the time of elevating $\Phi$ as the field operator. 

To determine the constant ${\cal N}$, we recall the inner product defined on the spacelike surface $\Sigma$, given by
\be
(\Phi_{\omega_1}, \Phi_{\omega_2}) = -i\int_{\Sigma} \sqrt{-\gamma}d^3x  ~n^{\mu} (\Phi_{\omega_1} \nabla_\mu\Phi_{\omega_2}^* - \Phi_{\omega_2}^* \nabla_\mu\Phi_{\omega_1}).
\ee
where $\Sigma$ is chosen to be a $\eta=const.$ hypersurface which also determines $n^{\mu} = 1/a (1,0,0,0)$. For the radiation dominated universe, we have $a = a_0 t^{1/2} = (a_0^2/2) \eta$ and, using the mode functions in \eqref{clphi} we get (after using $a_0=\sqrt{2{\cal H}e}$)
\be
{\cal N} = \frac{\sqrt{\omega}}{\sqrt{4\pi}{\cal H}e}.
\ee
Plugging this in the field expansion \eqref{clphi} and then changing the summation to integration we get 
\be 
\Phi = \int {d^3\omega} \left( u_{\omega}  a_\omega^{L} + v_{\omega}  a_\omega^{R} \right) + \text{h.c}
\ee
where
\ber
u_\omega = \frac{e^{-i\omega(\eta - r)}}{4\pi {\cal H}e\sqrt{\omega}~ \eta r} \label{mod1} \\
v_{\omega} = \frac{e^{-i\omega(\eta + r)}}{4\pi {\cal H}e \sqrt{\omega}~ \eta r} \label{mod2}
\eer
Now $\Phi$ can  be interpreted as the field operator which then defines the conformal vacuum for incoming and outgoing modes as $a_{\omega}^{L} |0\rangle_{(\eta)}^{L} =0$ and $a_{\omega}^{R} |0\rangle_{(\eta)}^{R} =0$.


\subsection{Quantum fields in spherically symmetric FRW}
\label{subh}
\subsubsection{Region-I (sub-Hubble)}
\label{subh1}
First we consider the massless scalar field equation in the background of metric (\ref{mRT2}) and look for a spherical wave type solution 
\begin{eqnarray}
\Phi_{\Omega}=\displaystyle{\sum_{l,m}}\frac{\Phi_{\Omega}^{lm}(T, R)}{R}Y_{lm}(\theta,u).
\end{eqnarray}
The $(R,T)$ dependent part decouples from the angular part (which is comprised of just the spherical harmonics) in the following way
\begin{eqnarray}
\left(\frac{\partial^2 \Phi_{\Omega}^{lm} }{\partial T^2}- \frac{\partial^2 \Phi_{\Omega}^{lm}}{\partial R^2} \right)+\frac{l(l+1)}{R(1-H^2R^2)}\Phi_{\Omega}^{lm} =0
\label{kgeq}
\end{eqnarray}
where we have made use of the properties of spherical harmonics $Y_{lm}(\theta,\phi)$. This equation shows that the $l\ne 0$ modes cannot reach the Hubble scale since the effective potential becomes infinite there. If we consider the $s-$wave modes then they are solutions of the equation
\begin{eqnarray}
\left(\frac{\partial^2 \Phi_{\Omega}^{00} }{\partial T^2}-\frac{\partial^2 \Phi_{\Omega}^{00} }{\partial R^2}\right) =0.
\label{l=0}
\end{eqnarray}
This equation is valid for all sub-huble modes $R\le 1/H$ (since the metric \eqref{mRT2} is valid only this region).
The mode solutions have the form
\begin{eqnarray}
\Phi_{\Omega}^{00} (T,R) \propto \exp{[-i\Omega (T \pm R)]}.
\end{eqnarray}
Now elevating $\Phi$ to the operator level, and introducing the new set of creation and annihilation operators $b_{\Omega},~b_{\Omega}^{\dagger}$ and using a continuous basis of mode functions we get
\be
\Phi = \int{{d^3\Omega} (b_{\Omega > \Omega_H}^{L} U_{\Omega}^{\text{sub}} + b_{\Omega > \Omega_H}^{R} V_{\Omega}^{\text{sub}}}) + \textrm{h.c.}
\label{sub}
\ee
where $\Omega > \Omega_H$ ensures that the modes are sub-Hubble and {\it ``L''}/{\it ``R''} once again stand for leftmoving/rightmoving modes with respect to an observer inside the Hubble radius. The mode functions are easily found to be 
\ber
U_{\Omega}^{\text{sub}} &=&  \frac{1}{4\pi \sqrt{\Omega} R} ~e^{-i\Omega (T-R)} \label{uo1}\\
V_{\Omega}^{\text{sub}} &=&  \frac{1}{4\pi \sqrt{\Omega} R}~ e^{-i\Omega (T+R)}.\label{vo1}
\eer
 The vacuum states for these two sectors are defined as $b_{\Omega >\Omega_H}^{L} | 0_{T}\rangle^{L} =0;~b_{\Omega >\Omega_H}^{R} | 0_{T}\rangle^{R}=0$. 


\subsubsection{Region-II (super-Hubble)}
Similarly, we can solve the wave equation in region-II in the background of the metric \eqref{mRT3} to find the super-Hubble modes. For that we use the ansatz
\begin{eqnarray}
\Phi=\displaystyle{\sum_{l,m,\Omega}}\frac{\Phi_{\Omega}^{lm}(T, R)}{T}Y_{lm}(\theta,u).
\end{eqnarray}
and under the $s-$wave approximation the field operator is expanded as
\be
\Phi = \int{\frac{d^3\Omega}{4\pi \sqrt{\Omega}} ({b}_{\Omega <\Omega_H}^{L} U_{\Omega}^{\text{sup}} + {b}_{\Omega <\Omega_H}^{R} V_{\Omega}^{\text{sup}}}) + \textrm{h.c.}
\label{sup}
\ee
and mode functions can be found for $T > R_H = 1/H$ as
\ber
U_{\Omega}^{\text{sup}} &=&  \frac{1}{T} ~e^{-i\Omega (T-R)} \label{uo2}\\
V_{\Omega}^{\text{sup}}&=&  \frac{1}{T}~ e^{-i\Omega (T+R)}. \label{vo2}
\eer


Now a word on completeness of the modes. For the s-waves, it is possible to obtain a complete set of orthogonal modes (which can be normalised in discrete basis), by considering both the sub-Hubble and super-Hubble modes, given in \eqref{sub} and \eqref{sup}. The modes in the background of \eqref{mRT2} has a vanishing support in the super-Hubble region, whereas, for \eqref{mRT3} this is true in the sub-Hubble  region. 

This, in turn, makes sure that there will be two sets of Bogolyubov coefficients, given by the sub and super Hubble modes, when we discuss particle creation for these new set of observers. This is formally different than the simpler two dimensional example discussed before.


\subsection{Bogolyubov coefficients and particle content}
We are interested in a specific case where the vacuum state $|0_{\eta}\rangle^{L/R}$ (``$L$'' stands for the left moving and ``$R$'' stands for the right moving modes) shows particle excitations with respect to the observers with proper time $T$. From our experience in black holes and Unruh radiation one expects $~^{L/R}\langle 0_{\eta}|b_\Omega^{\dagger L/R}b_\Omega^{L/R}|0_{\eta}\rangle^{L/R} \ne 0$ leading us to the nontrivial case of particle creation in the radiation phase of FRW universe. We shall calculate Bogolyubov coefficients for both left-moving and right-moving sectors and the particle content of these modes. The most relevant Bogolyubov coefficient is $\beta_{\Omega\omega}$ and this determines the average number of particles of frequency $\Omega$ by the relation
\ber
\langle n_{\Omega}\rangle^{L/R} &=& ~^{L/R}\langle 0_{\eta}|b_\Omega^{\dagger L/R}b_\Omega^{L/R}|0_{\eta}\rangle^{L/R} \nonumber \\
                                                                   & =& \int_{0}^{\infty} d\omega |\beta_{\Omega\omega}^{L/R}|^2 \label{no}
\eer

Once again, by looking at the trajectories of new observers in Fig. \ref{fig1} we expect the particle creation to take place in the regions where the observer is accelerating or decelerating. We also expect, given the symmetry of the trajectories in the sub and super-Hubble regions, for the particle content, to be also symmetric.

\subsubsection{Sub-Hubble modes ($\Omega > \Omega_{H}$)}

Here, we are interested in an observer inside the Hubble radius. The relevant mode functions, in the new coordinates, are calculated in subsection \eqref{subh1}. 
Considering the left moving modes (with $V=const.$) the Bogolyubov coefficient of our interest is
\be
\beta_{\Omega\omega}^{L} = 2i \int_{V=const.} dU R^2 d\tilde{\Omega} U_{\Omega} \partial_{U} u_{\omega}.\label{bog}
\ee  
where $d\tilde{\Omega}= \sin\theta d\theta d\phi$. By choosing $V = 0$ hypersurface for integration we have
\bwt
\be
\beta_{\Omega\omega}^{L} = -\frac{i}{2\pi \sqrt{\Omega\omega}} \int_{0}^{\infty} dU e^{-i(\Omega U + \omega\sqrt{2U/{\cal H}e})}\left(\frac{1}{U} + \frac{i \omega}{\sqrt{2{\cal H}eU}} \right).\label{int1}
\ee
\ewt
For simplicity, we perform the integration by restricting ourselves to the situation such that the ratio between frequencies in two frames satisfy the condition $\frac{\omega}{\Omega}<<{\cal H}e$ (cosmological value of ${\cal H}\sim 10^{37} s^{-1}$). The physical implication of this is already discussed before for the two dimensional case. The final expression is found to be{\footnote{Detailed calculation is showed in Appendix \ref{outbog}.}
\be
\beta_{\Omega\omega}^{L} = -\frac{ie^{\frac{i\omega^2}{2{\cal H}e\Omega}}}{2\pi\sqrt{\Omega\omega}} \left(1-\omega\sqrt{\frac{\pi}{4{\cal H}e \Omega}}  + i\omega\sqrt{\frac{\pi}{4{\cal H}e \Omega}}\right)
\label{betaout}
\ee 
Now substituting $|\beta_{\Omega\omega}^{L}|^2$ in \eqref{no} and integrating we find the particle excitation number density as
\be
\langle n_{\Omega} \rangle^{L} = \frac{1}{16\pi^3 \Omega} \left(\log{\omega} -\sqrt{\frac{\pi}{{\cal H}e\Omega}}\omega + \frac{\pi}{4{\cal H}e \Omega} \omega^2 \right).  \label{nout}
\ee
The logarithmic term shows an infra-red divergence as $\omega\rightarrow 0$ in $\langle n_{\Omega} \rangle^{out}$ just like the case of particle creation by moving mirror \cite{birrel}. In comparatively higher frequency $\omega^2$ term dominates and if we substitute $\frac{\omega}{\Omega} = \epsilon e {\cal H}$ (with $\epsilon<<1$), we get
\be
\langle n_{\Omega} \rangle^{L} \simeq \frac{{\cal H}e}{64\pi^2}\epsilon^2
\ee
which is again quite small and interestingly independent of $\Omega$, just like the two dimensional case. It is, however, capable to contribute a finite energy density if one counts all the particles created during the entire lifetime of radiation stage of the universe, as well as, considers all the frequencies within the above mentioned limit. Again, it is necessary to employ a fully numerical analysis to go beyond this limit and we do not consider this in the present article. 

For the right-moving modes (with $U=const.$), the Bogolyubov coefficient is
\be
\beta_{\Omega\omega}^{R} = 2i \int_{U=const.} dV R^2 d\tilde{\Omega} V_{\Omega} \partial_{V} v_{\omega}
\ee  
We chose $U=0$ hypersurface for performing the integration over $V$ where $0 < V < \infty$. After simplifying, the above expression becomes
\bwt
\be
\beta_{\Omega\omega}^{R} =  -\frac{i}{2\pi\sqrt{\Omega\omega}} \int_{0}^{\infty} dVe^{-i(\Omega V + \omega\sqrt{2V})}\left(\frac{1}{V} + \frac{i \omega}{\sqrt{2{\cal H}eV}} \right).
\label{boin}
\ee
\ewt
which is identical to \eqref{int1}, with $U$ replaced by $V$ and, therefore, leads to $\langle n_{\Omega} \rangle^{L} = \langle n_{\Omega} \rangle^{R}$ in \eqref{nout}.

\subsubsection{Super-Hubble modes ($\Omega < \Omega_{H}$)}
While calculating the Bogolyubov coefficients in region II (i.e., super-Hubble region) we should consider the mode functions those have non-vanishing support in that region. The mode functions in $(\eta,~r)$ coordinates have non-vanishing support both at the sub-Hubble and super-Hubble regions. However, as we have already shown, in $(T,~R)$ coordinates, it is only the mode functions \eqref{uo2} and \eqref{vo2}, that have non-vanishing support in the super-Hubble region.  Therefore they will be used to calculate the Bogolyubov coefficients in region II. 

For the left-moving sector we need to calculate
\be
\beta_{\Omega\omega}^{'~L} = 2 i\int_{V} dU T^2 d\tilde{\Omega} U_{\Omega} \partial_{U} u_{\omega}.
\ee
where the relevant mode functions are given in \eqref{mod1} and \eqref{uo2}. Note that now the radius of the two sphere is given by $T$ which, in region II, is expressed in \eqref{ter}. After simplifying, and again integrating over the $V =0$ axis, the integral becomes
\bwt
\be
\beta_{\Omega\omega}^{'~L} = -\frac{i}{2\pi\sqrt{\Omega\omega}} \int_{0}^{-\infty} dU e^{-i( - \Omega U + \omega\sqrt{-2U/{\cal H}e})}\left(\frac{1}{-U} + \frac{i \omega}{\sqrt{-2{\cal H}eU}} \right).\label{ints1}
\ee
\ewt
Notice that, in region II, we have $-\infty < U <0$ and the above integration is identical to \eqref{int1} upon the replacement of $U$ by $-U$. Therefore, we end up with the result \eqref{nout} for an average particle excitation for the super-Hubble modes for an observer in region II. It is also trivial to check that one would end up with an identical expression for particle excitation for the right-moving modes in region II. This fulfils our expectation that, particle content for sub and super-Hubble region is identical, simply because of the symmetry of the trajectories in these two regions, shown in Fig. \ref{fig1}.



\section{Discussion and future outlook}
We have discussed new geometric and field theoretic aspects of the radiation dominated era of the early universe. Specifically, we considered the radiation stage that joins the inflationary stage while extrapolated towards past. A new conformal transformation of the cosmological null coordinates was introduced which transformed the homogeneous, maximally symmetric FRW into a spherically symmetric, inhomogeneous spacetime during the radiation stage. To the best of our knowledge this is a new finding. 

While the  static observers  in the new frame do not see any horizon, some non-static observers encounter a horizon due to their motion. The later situation in our opinion has a nice analogy with the Rindler observers in Minkowski spacetime. 

We discussed the particle excitation of the cosmological vacuum state, for which, we  kept ourselves confined only to the static case. Static observers detect particles in the cosmological vacuum state. The trajectory of these observers is such that, it matches the freely falling observer trajectory in the infinite past, in the super-Hubble region.  Then the observers start accelerating and reaches the luminal velocity while crossing the Hubble horizon and decelerates in sub Hubble region to finally reach a free fall towards the centre, in distant future. Since there is an acceleration (and deceleration) involved, there should be particle creation and this is what we further discussed in this paper.

The symmetry of the new metrics, in sub and super Hubble regions, allows a new quantisation of the matter fields (massless scalar field with arbitrary coupling). The well known conformal vacuum was shown to have particle excitations with respect to the new set of observers with the new coordinates. The particle content of the conformal vacuum was computed both in two and four spacetime dimensions. 

As a future prospect one might also want to show the possible connection of this new coordinates with the static de Sitter coordinates in the inflationary stage. This will initiate a possibility of  exploring the field evolution from the de Sitter universe into this new spacetime. That in principle can have important consequences. This is an interesting problem and currently under further investigation \cite{modak}.  Another work that needs to be done is to understand the new physics related with the non-static observers which is totally kept open at this stage.

Perhaps, the most important outstanding issue is to clarify the relevance of these new coordinates introduced here, in describing some natural phenomena of the universe. From the static observers' perspective, since they are ever accelerating in the super-Hubble region, the time for muon decay becomes longer and longer, reaching a situation where muon does not decay at all (while crossing the Hubble scale) {\footnote{I thank Prof. Satoshi Iso for pointing this out.}}. Then in the sub Hubble region this phenomena takes place in a reverse order. Of course, any static charge in the cosmological frame does not remain static in this new frame and phenomena like Bremsstrahlung takes place with respect to static observers. Except these comments, however, at this point we are unable to pinpoint  if these new coordinates can help us understanding something fundamental about our universe. We shall investigate further this possibility in future.

\section{Acknowledgements}
I thank Prof. T. Padmanabhan for several discussions which led me to this problem and his key inputs in the early stages of the work. I cannot thank enough Prof. Satoshi Iso for his key conceptual and technical inputs, which have considerably improved the manuscript.  I want to thank the anonymous referee for his/her valuable inputs which helped me to introduce the discussion of non-static observers. Part of my research was carried out at KEK, when I was an International Research Fellow of Japan Society for Promotion of Science (JSPS). This research is also supported by a start up research grant from PRODEP-SEP, M\'exico.
 

\appendix

\section{Calculation of Bogolyubov coefficient \eqref{nout}}
\label{outbog}
First using the mode function \eqref{uo1} and the relationship between the coordinates \eqref{Uu-n} we obtain
\be
\partial_U u_{\omega} = \frac{e^{-i\omega u}}{4\pi \sqrt{\omega}R}\left(\frac{1}{V-U} -  \frac{i\omega}{\sqrt{2{\cal H}e U}} \right)
\ee
Substituting $U_\Omega$ from \eqref{uo1} and the above expression, and integrating over the angular part in \eqref{bog} we get  \eqref{int1}. Next we substitute $U$ by $u$ using the relationship \eqref{Uu-n} in \eqref{int1}
\be
\beta_{\Omega\omega} = -\frac{i}{2\pi\sqrt{\Omega\omega}} \int_0^{\infty} du e^{-i\omega u - i\Omega {\cal H} e u^2/2}(i\omega + 2/u)
\ee

\be
\beta_{\Omega\omega} = -\frac{ie^{\omega^2/2{\cal H} e\Omega}}{2\pi\sqrt{\Omega\omega}} (i\omega I_{1} + 2I_{2}) \label{bogn}
\ee

\ber
I_{1} &=& \int_0^{\infty} du e^{\frac{-i\Omega{\cal H} e}{2}(u+\frac{\omega}{{\cal H} e\Omega})^2}\\
I_{2} &=& \int_0^{\infty} \frac{du}{u} e^{\frac{-i\Omega{\cal H} e}{2}(u+\frac{\omega}{{\cal H} e\Omega})^2}
\eer
Now we make a change of variable by setting $x=(u+\frac{\omega}{{\cal H} e\Omega})^2$. This imply
\ber
I_{1} &=& \int_{x_0}^{\infty}\frac{ dx }{2\sqrt{x}}e^{\frac{-i\Omega{\cal H} e}{2}x}\\
I_{2} &=&  \int_{x_0}^{\infty}\frac{ dx }{2\sqrt{x}(\sqrt{x} -\sqrt{x_0})}e^{\frac{-i\Omega{\cal H} e}{2}x}\eer
where $x_0 = (\frac{\omega}{{\cal H} e\Omega})^2$. In the limit $\omega/\Omega << {\cal H}$ we have $x_0 \rightarrow 0$ and using the identity \eqref{id} it is easy to check
\ber
\lim_{x_0\rightarrow 0} I_1 &=& e^{i\pi/4} \sqrt{\frac{\pi}{2{\cal H}e \Omega}}\\
\lim_{x_0\rightarrow 0} I_2 &=& 1/2.
\eer
Finally substituting these expressions in \eqref{bogn} we find \eqref{nout}.

\end{document}